\documentclass[conference]{IEEEtran}

\usepackage{amsmath,amssymb,amsthm}
\usepackage{enumerate}
\usepackage{stfloats}
\usepackage{comment}
\usepackage{subcaption}
\usepackage{siunitx}
\usepackage{mathtools}
\usepackage{accents,latexsym,cancel}
\usepackage{cite}

\usepackage[dvipsnames]{xcolor}
\definecolor{myPink}{RGB}{255,105,183}

\usepackage[T1]{fontenc}
\usepackage{graphics} 
\usepackage{epsfig} 
\usepackage[mathscr]{euscript}
\usepackage{algorithm}
\usepackage[noend]{algpseudocode}
\usepackage{bbm}
\makeatletter
\def\BState{\State\hskip-\ALG@thistlm}
\makeatother

\usepackage{tikz}
\usetikzlibrary{arrows,shapes,chains,matrix,positioning,scopes,patterns,calc,
decorations.markings,
decorations.pathmorphing,
}

\usepackage{pgfplots}
\pgfplotsset{compat=1.3}
\usepgflibrary{shapes}

\renewcommand{\epsilon}{\varepsilon}

\newcommand{\RNum}[1]{\uppercase\expandafter{\romannumeral #1\relax}}

\newcommand{\gv}{\ensuremath{\mathbf{g}}}
\newcommand{\mv}{\ensuremath{\mathbf{m}}}
\newcommand{\Lv}{\ensuremath{\mathbf{L}}}
\newcommand{\pv}{\ensuremath{\mathbf{p}}}

\newcommand{\rv}{\ensuremath{\mathbf{r}}}
\newcommand{\sv}{\ensuremath{\mathbf{s}}}
\newcommand{\vv}{\ensuremath{\mathbf{v}}}

\newcommand{\wv}{\ensuremath{\mathbf{w}}}
\newcommand{\xv}{\ensuremath{\mathbf{x}}}
\newcommand{\yv}{\ensuremath{\mathbf{y}}}
\newcommand{\zv}{\ensuremath{\mathbf{z}}}

\newcommand{\Ktot}{\ensuremath{K_{\mathrm{tot}}}}
\newcommand{\Ka}{\ensuremath{K_{\mathrm{a}}}}

\def\Pr{\mathrm{Pr}}

\DeclareMathAlphabet{\mcl}{OMS}{cmsy}{m}{n}

\newlength\tikzwidth
\newlength\tikzheight

\definecolor{mycolor1}{rgb}{0.63529,0.07843,0.18431}%
\definecolor{mycolor2}{rgb}{0.00000,0.44706,0.74118}%
\definecolor{mycolor3}{rgb}{0.00000,0.49804,0.00000}%
\definecolor{mycolor4}{rgb}{0.87059,0.49020,0.00000}%
\definecolor{mycolor5}{rgb}{0.00000,0.44700,0.74100}%
\definecolor{mycolor6}{rgb}{0.74902,0.00000,0.74902}%

\newif\ifproof
\prooftrue

\def\fig_path{./Figures}
\IEEEoverridecommandlockouts
\begin{document}
\title{On Approximate Message Passing for Unsourced Access with Coded Compressed Sensing
}

%
\author{\dag Vamsi K. Amalladinne, \dag Asit Kumar Pradhan,  \S Cynthia Rush,
\dag Jean-Francois Chamberland, \dag Krishna R. Narayanan\\
 \dag Department of Electrical and Computer Engineering, Texas A\&M University\\
 \S Department of Statistics, Columbia University
\thanks{
This material is based upon work supported, in part, by the National Science Foundation (NSF) under Grant No.~CCF-1619085 and by Qualcomm Technologies, Inc., through their University Relations Program.}}

\maketitle

\begin{abstract}
Sparse regression codes with approximate message passing (AMP) decoding have gained much attention in recent times.
The concepts underlying this coding scheme extend to unsourced access with coded compressed sensing (CCS), as first pointed out by Fengler, Jung, and Caire.
More specifically, their approach uses a concatenated coding framework with an inner AMP decoder followed by an outer tree decoder.
In the original implementation, these two components work independently of each other, with the tree decoder acting on the static output of the AMP decoder.
This article introduces a novel framework where the inner AMP decoder and the outer tree decoder operate in tandem, dynamically passing information back and forth to take full advantage of the underlying CCS structure.
The enhanced architecture exhibits significant performance benefit over a range of system parameters.
Simulation results are provided to demonstrate the performance benefit offered by the proposed access scheme over existing schemes in the literature.
\end{abstract}


\section{Introduction and Background}
\label{section:Introduction}
Unsourced random access is a novel communication paradigm envisioned to accommodate the increasing traffic demands of next generation wireless networks.
This framework garnered significant research interest owing to the emergence of Internet of Things (IoT) and machine-driven communications.
This model differs from the conventional multiple access paradigm in a number of ways.
Conventional multiple access schemes are more suited for human-centric communications with sustained connections, where the cost of coordination can be amortized over a long time period.
However, this may not be possible in machine-centric communications where device transmissions are sporadic with very short payloads.
This new reality invites the design of a protocol in which it is not mandatory for active devices to reveal their identities.
Rather, decoding is done only up to a permutation of the transmitted payloads, without regard for the identities of transmitting devices.
Active devices who wish to reveal their identities can embed this information in their payloads.
This approach enables all the active devices to share a common codebook for their transmissions.

A random coding achievability bound for the unsourced random access channel in the absence of complexity constraints is derived in \cite{polyanskiy2017perspective}.
Subsequently, a number of practical coding schemes that aim to perform close to this conceptual benchmark have been proposed in the literature \cite{ordentlich2017low,vem2019user,amalladinne2019coded,Giuseppe,pradhan2019joint,calderbank2018chirrup,marshakov2019polar,pradhan2019polar}. 
In this line of work, Amalladinne et al.~\cite{amalladinne2019coded} put forth a concatenated coding scheme that uses an inner compressed sensing (CS) code and an outer tree code.
They take advantage of the connection between unsourced multiple access and support recovery in high dimensional compressed sensing.
A divide-and-conquer approach is leveraged to split the information messages of active users into several sub-blocks, each amenable to standard CS solvers. Redundancy is employed in the form of an outer tree code to bind the information sub-blocks that correspond to one message together.
This scheme employs the parity-check bits added in the encoding phase solely for the purpose of stitching.
Yet, it turns out that the inner and outer decoders in CCS can be executed in tandem, and the redundancy employed during the transmission phase can be utilized to curtail the realm of possibilities for parity-check bits in subsequent stages \cite{amalladinne2019enhanced}.
This algorithmic improvement developed for CCS offers significant benefits both in terms of error performance and computational complexity.
The impetus for our research is the hope that similar notions may apply to other schemes related to unsourced random access.

Recall that approximate message passing (AMP) refers to a broad class of iterative algorithms derived from message passing algorithms on dense factor graphs.
For instance, AMP has been successfully applied to the problem of reconstructing sparse signals from a small number of noisy measurements.
The first application of an AMP decoder to the unsourced MAC problem is due to Fengler, Jung, and Caire~\cite{Giuseppe}.
Therein, the authors draw a connection between the structure of coded compressed sensing (CCS)~\cite{amalladinne2019coded} and sparse regression code (SPARC) constructions~\cite{CIT-092}.
They then extend the CCS framework~\cite{amalladinne2018couple} by using a design (measurement) matrix that does not assume a block diagonal structure, and they apply AMP as part of the decoding process.
In this article, we leverage the insights developed in \cite{amalladinne2019enhanced} to devise novel message passing rules that integrate the tree code and AMP.
We show that the parity-check bits employed in tree code can be used to assist the convergence of AMP.
We also provide finite-block length numerical results for this proposed scheme to demonstrate the performance gain it offers over the scheme in \cite{Giuseppe}.

\section{System Model} 
\label{section:SystemModel}
Consider a situation where $\Ka$ active devices out of a total of $\Ktot$ such devices each wish to send a message to an access point.
The transmission process takes place over a multiple access channel, with a time duration of $n$ channel uses (real degrees of freedom).
The signal received at the destination is given by
\begin{equation} \label{equation:ChannelModel}
\textstyle \yv = \sum_{i=1}^{\Ka} \xv_i + \zv
\end{equation}
where $\xv_i \in \mathcal{C} \subset \mathbb{R}^n$ is the codeword transmitted by active device $i$. 
The noise component $\zv$ is composed of an independent sequence of Gaussian elements, each with distribution $\mathcal{N}(0,1)$.
The devices share a common codebook and, as such, $\xv_i$ is a function of the payload of device~$i$, but not of its identity.
The selection of a codeword follows the general structure obtained by combining the tree code of Amalladinne et al.~\cite{amalladinne2018couple} and the SPARC-like encoding of Fengler et al.~\cite{Giuseppe}.
While the broad structure of the outer code is inspired by \cite{amalladinne2018couple}, we make key modifications that allow us to pass messages back and forth between blocks during AMP decoding in an efficient manner, the details of which will be clear in subsequent sections.

Consider a payload $\wv \in \{0, 1\}^w$.
Redundancy is first added to this message in the form of a tree code.
That is,  this $w$-bit binary message $\wv$ is enhanced with $p$ parity-check bits and the coded block of length $v=w+p$ is partitioned into  $L$ sub-blocks with lengths $v_1, v_2, \ldots, v_L$ such that $\sum_{\ell=1}^L v_{\ell} = v$.
Overall, the organization of the encoded message assumes the form $\vv = {\vv(1)}  {\vv(2)} \cdots {\vv(L)},$ where $\vv(\ell)$ denotes the $\ell$th sub-block.
The tree code proposed in \cite{amalladinne2018couple} features sub-blocks containing a combination of information and parity-check bits.
However, in our current treatment, each coded sub-block features either information bits or parity-check bits, but not a combination of both.
While not obvious, this eliminates dependencies among sub-blocks and allows us to exploit a circular convolution structure propitious to FFT.
We denote the collection of information sub-blocks by $\mathcal{W}$ and the parity sub-blocks by $\mathcal{P} = [1:L] \setminus \mathcal{W}$.
The parity-check bits contained in a sub-block act as constraints on information bits corresponding to a subset of $\mathcal{W}$.
As part of the next encoding step, each sub-block is turned into a message index of length $m_{\ell} = 2^{v_{\ell}}$.
This action is a cornerstone of CCS and, consequently, it is worth going over details carefully.
Mathematically, we have
\begin{equation} \label{equation:IndexFunction}
\begin{split}
\mv(\ell) &= f_{\mathbb{F}_2^{v_{\ell}} \rightarrow \{ 0, 1 \}^{m_{\ell}}} (\vv(\ell)) 
\end{split}
\end{equation}
where the function $ f_{\mathbb{F}_2^{v_{\ell}} \rightarrow \{ 0, 1 \}^{m_{\ell}}}$ can be described by regarding argument $\vv(\ell)$ as an integer in binary form.
The output is then a length-$2^{v_{\ell}}$ real vector with zeros everywhere, except for a one at location $[\vv(\ell)]_2$, where
the shorthand notation $[\cdot]_2$ stands for an integer expressed with a radix of 2.

A block-sparse message $\mv$ is subsequently created by concatenating individual sections, $\mv = \mv(1) \mv(2) \cdots \mv(L)$.
We describe vector $\mv$ as block-sparse because every section $\mv(\ell)$ features exactly one non-zero element.
The induced vector is reminiscent of a sparse regression code~\cite{joseph2013fast,CIT-092}.
We emphasize that the structure above is slightly more general than the form in~\cite{Giuseppe} since it accommodates the possibility of having blocks of different sizes; the resemblance is nevertheless manifest.

Let $\mathbf{A}$ be an $n \times m$ matrix over the real numbers, where $m = \sum_{\ell=1}^L m_{\ell}$.
Transmitted signals in $\mathcal{C}$ are obtained via the product $\xv = \mathbf{A} \mv$ over the field of real numbers.
With all the active devices utilizing the same codebook, this process yields a received vector $\yv$ of the form
\begin{equation} \label{equation:SPARClike}
\textstyle
\yv = \sum_{i=1}^{\Ka} \mathbf{A} \mv_i + \zv
= \mathbf{A} \left( \sum_{i=1}^{\Ka} \mv_i \right) + \zv
= \mathbf{A} \sv + \zv
\end{equation}
where $\sv = \sum_{i=1}^{\Ka} \mv_i$.
The resulting multiple access channel can be viewed as the combination of a point-to-point channel $\sv \rightarrow \mathbf{A} \sv + \zv$ and an outer binary adder MAC $\sv = \sum_{i=1}^{\Ka} \mv_i$.
The authors in \cite{Giuseppe} refer to these components as the \emph{inner} and \emph{outer channels}, respectively.
They also draw a distinction between the \emph{inner} and \emph{outer encoder/decoder} pairs.

While we embrace the aforementioned categorization for the channel components, \emph{inner} and \emph{outer channels}, we do not subscribe to the latter dissociation between the decoders.
Rather, we seek to exploit the fact that decoding can be improved by allowing information to flow dynamically between the inner and outer components while decoding takes place.
As mentioned above, the impetus behind this novel perspective stems from a potential algorithmic enhancement that was first noticed in the context of coded compressive sensing~\cite{amalladinne2019enhanced}.

Before turning to a detailed exposition of our proposed framework, we briefly review the evaluation criterion that underlies much of the past contributions pertaining to the unsourced MAC~\cite{polyanskiy2017perspective}.
Based on the received signal $\yv$, the access point is tasked with producing a list $\widehat{W}(\yv)$ of estimated messages.
The size of the list should not exceed $\Ka$, i.e., $| \widehat{W}(\yv) | \le \Ka$.
Performance is assessed based on the per-user probability of error,
\begin{equation} \label{equation:PUPE}
\textstyle
P_{\mathrm{e}} = \frac{1}{\Ka} \sum_{i = 1}^{\Ka}
\Pr \left( \wv_i \notin \widehat{W}(\yv) \right) .
\end{equation}
We are now ready to initiate our treatment of CCS-AMP.

\section{Proposed Coding Scheme}
\label{sectionProposedCodingScheme}

In this section, we begin with the description of the outer tree code, which necessitates some modifications tailored to its integration with the AMP framework.
We then proceed with the enhanced AMP inner code.

\subsection{Tree Encoding:}
\label{TreeEncoder}

This section focuses on the \emph{outer code}, which takes the form of a modified tree code.
An important distinction between the original CCS algorithm and CCS-AMP from a tree code perspective stems from the fact that the former is applied to short lists (approximately) of size $\Ka$, whereas AMP produces long lists of sub-blocks and likelihoods as the decoder goes through iterations.
To accommodate this situation, the tree code employed in this article differs slightly from the original tree code introduced in~\cite{amalladinne2018couple,amalladinne2019coded}.
In a manner akin to the original CCS scheme, the construction of our tree code starts by splitting message bits into fragments.
Parity patterns are then added to fragments in a causal fashion, leading to vector $\vv$.
To this extent, the alternate tree code subscribes to the same structure as the original one found in~\cite{amalladinne2018couple}.
However, in our revised construction, parity bits are created differently.
For $\ell \in \mathcal{P}$, we denote the collection of information sub-blocks on which the parity block $\vv(\ell)$ acts by $\mathcal{W}_{\ell} \subset \mathcal{W}$.
The parity block $\vv(\ell)$ is obtained using the following three-step sequence.
We first take a random linear combination of the information bits in each fragment of $\mathcal{W}_{\ell}$; that is, $\vv(j) \mathbf{G}_{j,\ell}$ for $j \in \mathcal{W}_{\ell}$.
Within this step, vector operations are taken over Galois field $\mathbb{F}_2$.
We transpose these combinations from the space of length-$v_{\ell}$ binary vectors over $\mathbb{F}_2$ to the ring of integers modulo-$2^{v_{\ell}}$, which we denote by $\mathbb{Z}/2^{v_{\ell}}\mathbb{Z}$.
We then add the resulting elements of $\mathbb{Z}/2^{v_{\ell}}\mathbb{Z}$ using modulo-$2^{v_{\ell}}$ arithmetic.
Finally, we convert the ensuing sum back to a vector in $\mathbb{F}_2^{v_{\ell}}$.
The resulting sequence of bits serves as the parity sub-block $\vv(\ell)$.
Mathematically, these operations can be expressed as
\begin{align}
\textstyle
\vv(\ell)
&\equiv \sum_{j \in \mathcal{W}_{\ell}}
\big[ \vv(j) \mathbf{G}_{j,\ell} \big]_{\mathbb{Z}/2^{v_{\ell}}\mathbb{Z}} \label{equation:ParityGeneration}
\end{align}
Above, the notation $[ \cdot ]_{\mathbb{Z}/2^{v_{\ell}}\mathbb{Z}}$ emphasizes that the argument is interpreted as an element of the quotient ring $\mathbb{Z}/2^{v_{\ell}}\mathbb{Z}$, and the equivalence relation `$\equiv$' denotes equality in $\mathbb{Z}/2^{v_{\ell}}\mathbb{Z}$.

The entries in $\mathbf{G}_{j,\ell} \in \{0,1\}^{v_j \times v_l}$ are independent Bernoulli trials, each with parameter half.
Conceptually, $\vv(\ell)$ contains non-linear constraints on the information bits from fragments in $\mathcal{W}_{\ell}$.
It is worth mentioning that the parity encoding process in \eqref{equation:ParityGeneration} is more contrived than the random linear combinations utilized in the original tree code~\cite{amalladinne2018couple}.
The value of this somewhat intricate encoding process is that it
induces a cyclic structure on parity precursors $\{ \vv(j) \mathbf{G}_{j,\ell} \}$ conducive to circular convolution, which is befitting to the eventual application of FFT techniques.

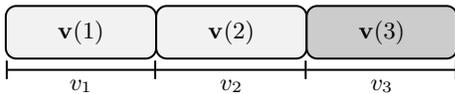
\begin{figure}[htb]
  \centering
  \begin{tikzpicture}[
  font=\small, >=stealth', line width=0.75pt,
  infobits/.style={rectangle, minimum height=7mm, minimum width=20mm, draw=black, fill=gray!10, rounded corners},
  paritybits/.style={rectangle, minimum height=7mm, minimum width=20mm, draw=black, fill=gray!40, rounded corners}
]

\node[infobits] (vb1) at (1,0) {$\vv(1)$};
\node[infobits] (vb2) at (3,0) {$\vv(2)$};
\node[paritybits] (vp3) at (5,0) {$\vv(3)$};
\draw[|-|] (0,-0.5) to node[midway,below] {$v_1$} (2,-0.5);
\draw[-|] (2,-0.5) to node[midway,below] {$v_2$} (4,-0.5);
\draw[-|] (4,-0.5) to node[midway,below] {$v_3$} (6,-0.5);

\end{tikzpicture}
  \caption{An information and parity allocation that is conducive to the application of FFT-based techniques appears above.
  Sub-blocks are homogeneously composed of information or parity bits, but not both.
  Shaded sub-block denotes a parity fragment.}
  \label{figure:subvectors2}
\end{figure}

\subsection{Tree Decoding}
\label{subsection:TreeDecoding}

This section focuses on the (soft) decoding of the \emph{outer tree code}, which can be employed in tandem with the iterative decoding of the \emph{inner code}.
The conceptual starting point for this discussion is a collection of functions, one for each section, that capture the likelihoods of the corresponding sub-blocks.
For instance, $\mathcal{L}_1 (\cdot)$ can output the likelihood of any root fragment;
whereas $\mathcal{L}_{\ell}(\cdot)$ takes as argument a sub-block at level~$\ell$ and returns its likelihood.

The encoding process and, specifically, the parity generation defined in \eqref{equation:ParityGeneration} induce a generalized Markov structure on the tree code.
Given likelihoods at every stage, the posterior probabilities of sub-blocks can be computed using the forward-backward algorithm.
Unfortunately, for the problem at hand, the complexity of implementing this optimal algorithm can be cost prohibitive.
Consequently, we focus on certain aspects of the coded structure, computing prior probabilities for $\vv(\ell)$ conditioned on pertinent observations and using the parity generation equation of \eqref{equation:ParityGeneration}.

For notational convenience, we use $\vv(\mathcal{A})$ to represent the collection of sub-blocks $\{\vv(j): j \in \mathcal{A} \subset [1:L]\}$.
Mathematically, for all $\ell \in \mathcal{P}$, the probability of truncated sequence $\{\hat{\vv}(\mathcal{W}_{\ell}), \hat{\vv}(\ell)\}$ conditioned on the likelihoods of the fragments in $\mathcal{W}_{\ell}$ can be written as
\begin{equation} \label{equation:TreeDecoderParitalMAP}
\begin{split}
&\Pr \left( \hat{\vv}(\mathcal{W}_{\ell}), \hat{\vv}(\ell) |  \left\{\mathcal{L}_{j} \left( \hat{\vv}(j) \right): j \in \mathcal{W}_{\ell} \right\} \right)\\
&\propto \mathcal{G} \left(  \hat{\vv}(\mathcal{W}_{\ell}), \hat{\vv}(\ell) \right)
\prod_{j \in \mathcal{W}_{\ell}} \mathcal{L}_{j} \left( \hat{\vv}(j) \right) ,
\end{split}
\end{equation}
where $\mathcal{G} (\cdot)$ is an indicator function that returns one when the sub-blocks in its argument are parity consistent with tree encoding, and it returns zero otherwise.
A more relevant quantity for the joint decoding of the \emph{inner} and \emph{outer codes} is the probability of $\hat{\vv}(\ell)$ conditioned on the likelihoods of fragments in $\mathcal{W}_{\ell}$.
For a single user, we write this quantity as
\begin{equation} \label{equation:TreeDecoderPosteriorV}
\begin{split}
&\Pr \left( \hat{\vv}(\ell) | \left\{\mathcal{L}_{j} \left( \hat{\vv}(j) \right): j \in \mathcal{W}_{\ell} \right\} \right) \\
&= \frac{ \sum_{\hat{\vv}(\mathcal{W}_{\ell})} \Pr \left( \hat{\vv}(\mathcal{W}_{\ell}), \hat{\vv}(\ell) \right)
\prod_{j \in \mathcal{W}_{\ell}}\mathcal{L}_j \left( \hat{\vv}(j) \right) }
{ \sum_{\hat{\vv}(\mathcal{W}_{\ell})} \Pr \left( \hat{\vv}(\mathcal{W}_{\ell}) \right)
\prod_{j \in \mathcal{W}_{\ell}}\mathcal{L}_j \left( \hat{\vv}(j) \right) } \\
&= \frac{ \sum_{\hat{\vv}(\mathcal{W}_{\ell})}
\mathcal{G} \left( \hat{\vv}(\mathcal{W}_{\ell}), \hat{\vv}(\ell) \right)
\prod_{j \in \mathcal{W}_{\ell}} \mathcal{L}_{j} \left( \hat{\vv}(j) \right)}
{\sum_{\hat{\vv}(\mathcal{W}_{\ell})}
\prod_{j \in \mathcal{W}_{\ell}} \mathcal{L}_{j} \left( \hat{\vv}(j) \right)} \\
&\propto  \sum_{\hat{\vv}(\mathcal{W}_{\ell})}
\mathcal{G} \left( \hat{\vv}(\mathcal{W}_{\ell}), \hat{\vv}(\ell) \right)
\prod_{j \in \mathcal{W}_{\ell}} \mathcal{L}_{j} \left( \hat{\vv}(j) \right).
\end{split}
\end{equation}
In words, the conditional probability of sub-block $\hat{\vv}(\ell)$, given $\left\{ \mathcal{L}_{j} \left( \hat{\vv}(j) \right) : j \in \mathcal{W}_{\ell} \right\}$, is proportional to the sum of the (joint) likelihoods of all the truncated paths $\hat{\vv}(\mathcal{W}_{\ell})$ that are parity consistent with $\hat{\vv}(\ell)$.
In some sense, the aggregate posterior of the parity consistent paths acts as a prior for the estimate of $\hat{\vv}(\ell)$.
This is the insight we seek to leverage in the overall decoding process.
The soft information of \eqref{equation:TreeDecoderPosteriorV} can be integrated in the overall decoding process and guide its convergence.

A naive implementation of \eqref{equation:TreeDecoderPosteriorV} would have the decoder look at $\prod_{j \in \mathcal{W}_{\ell}} v_{j}$ distinct paths to compute the prior on $\hat{\vv}(\ell)$, which is intractable for parameters of interest.
Our goal, then, is to exploit the structure of \eqref{equation:TreeDecoderPosteriorV} in getting low-complexity solutions.
We initiate this exercise by revisiting the last line of \eqref{equation:TreeDecoderPosteriorV},
\begin{align}
&\tilde{q}_{\ell} \left( \hat{\vv} (\ell) \right)
\propto \sum_{\hat{\vv}(\mathcal{W}_{\ell})}
\mathcal{G} \left( \hat{\vv}(\mathcal{W}_{\ell}), \hat{\vv}(\ell) \right)
\prod_{j \in \mathcal{W}_{\ell}} \mathcal{L}_{j} \left( \hat{\vv}(j) \right) \nonumber \\
&= \sum_{\hat{\vv}(\mathcal{W}_{\ell})}
\mathbf{1} \left( \sum_{j \in \mathcal{W}_{\ell}}
\left[ \hat{\vv}(j) \mathbf{G}_{j,\ell} \right]_{\mathbb{Z}/2^{v_{\ell}}\mathbb{Z}}
\equiv \hat{\vv}(\ell) \right) 
\prod_{j \in \mathcal{W}_{\ell}} \mathcal{L}_j \left( \hat{\vv}(j) \right) \nonumber \\
 &= \underbrace{\sum_{\substack{g_j \in \mathbb{Z}/2^{v_{\ell}}\mathbb{Z} \\
\sum_{j \in \mathcal{W}_{\ell}} g_j \equiv \hat{\vv}(\ell)}} \left( \prod_{j \in \mathcal{W}_{\ell}}
\left( \sum_{ \left[ \hat{\vv}(j) \mathbf{G}_{j,\ell} \right]_{\mathbb{Z}/2^{v_{\ell}}\mathbb{Z}} \equiv g_j}   \mathcal{L}_{j} \left( \hat{\vv}(j)
\right) \right) \right) }_{\text{circular convolution structure}}. \label{circConv1}
\end{align}
The circular discrete convolution structure identified above invites the immediate application of the discrete Fourier transform and related techniques.
In addition, since the underlying period is $2^{v_{\ell}}$ (a factor of two), these operations can be performed with the fast Fourier transform algorithm.
Thus, through the structure of the parity patterns produced by \eqref{equation:ParityGeneration}, the computation of soft posterior probabilities on paths becomes manageable under \eqref{circConv1}, even for large values of $2^{v_{\ell}}$.
This fact alone forms the impetus behind the development of an alternate tree code and the adoption of its more intricate parity generation process.

Procedurally, the probabilities for binary sequences of the form $\hat{\vv}(\ell):  \ell \in \mathcal{P}$ accounting for priors, likelihoods, and the parity structure of $\hat{\vv}(\ell)$ are collectively computed as follows.
At level~$j \in \mathcal{W}_{\ell}$, a static binary vector $\gv_{j,\ell}^{(g)} \in \{0, 1\}^{v_j}$ is stored for every $g \in \mathbb{Z}/2^{v_{\ell}}\mathbb{Z}$.
This vector features a one at every index location where $\hat{\vv}(j)$ is such that $\left[ \hat{\vv}(j) \mathbf{G}_{j,\ell} \right]_{\mathbb{Z}/2^{v_{\ell}}\mathbb{Z}} \equiv g$, and zeros everywhere else.
Through this partitioning, we get
\begin{equation} \label{equation:LvComponents}
\sum_{ \left[ \hat{\vv}(j) \mathbf{G}_{j.\ell} \right]_{\mathbb{Z}/2^{v_{\ell}}\mathbb{Z}} \equiv g}
 \mathcal{L}_{j} \left( \hat{\vv}(j)\right)
= \left\langle \left[ \mathcal{L}_{j} \left( \hat{\vv}(j) \right) \right]_{\mathbb{R}^{2^{v_j}}}, \gv_{j, \ell}^{(g)} \right\rangle .
\end{equation}
When ordered and stacked, the values in \eqref{equation:LvComponents} yield a vector $\Lv_{j,\ell}$ in $\mathbb{R}^{2^{v_{\ell}}}$.
In view of the circular convolution structure identified above, the ordered and stacked vector of $\mathbf{\tilde{q}}_{\ell}$ can be computed at once as
\begin{equation} \label{equation:BlockFFT}
\mathbf{\tilde{q}}_{\ell}  
= \frac{  \operatorname{FFT}^{-1} \left( \prod_{j \in \mathcal{W}_{\ell}} \operatorname{FFT} \left( \Lv_{j,\ell} \right) \right) }
{ \left\| \operatorname{FFT}^{-1} \left( \prod_{j \in \mathcal{W}_{\ell}} \operatorname{FFT} \left( \Lv_{j,\ell} \right) \right) \right\|_1 }.
\end{equation}
The denominator of the above expression acts as a normalization factor.
\subsubsection{Multi-Phase Decoding with Extended Lists}
In a hard version of tree decoding \cite{amalladinne2019coded}, lists at every stages contain $\Ka$ entries (or slightly more).
The tools described above and, in particular, \eqref{circConv1} admits the propagation of likelihoods over much longer lists.
For the parameters of interest, we are carry this process over at most three or four sub-blocks.

In view of these comments, our proposed approach is to have two or three sections of information bits followed by a section of parity bits that acts on these information sections, as depicted in Fig~\ref{figure:subvectors2}.
We can then run \eqref{circConv1} on these first few sub-blocks up to the first parity section.
Using \eqref{equation:BlockFFT}, the probabilities of individual sub-blocks at a particular stage can be updated using the likelihoods of elements at neighboring stages.
This step is explicit in \eqref{circConv1} for parity sections, but can be performed in an analogous manner for any stage of a circular convolution interval.

With posterior probabilities of individual sub-blocks in hand, lists at these stages can be pruned with high statistical confidence, discarding the least probable sub-blocks.
Parity constraints are then enforced on combinations of the surviving members, resulting in a reduced list of most likely super-sections.
The resulting stitched blocks then act as a fused section for the next sequence of information sub-blocks.
Selected super-sections are the equivalent of active paths in CCS~\cite{amalladinne2019coded}, except that the FFT structure enables the decoder to maintain many more entries and every partial paths has a likelihood associated with it.
We note that the parity bits employed at previous stages can be disregarded because their selective power is already accounted for in the super-section.
We can carry this process forward, using blocks of parity to prune and bind information sections at various points.
Repeated applications of these concepts can be scaffolded in a cascading, hierarchical, or mixed fashion.
Having gained the ability to compute probabilities of the form $\tilde{q}_{\ell} \left( \hat{\vv}(\ell) \right)$ efficiently and to subsequently prune and stitch paths, we are ready to initiate our description of the CCS-AMP algorithm.
\subsection{Inner Code and AMP Decoding}
As mentioned in Section~\ref{section:SystemModel}, the inner code introduced in \eqref{equation:SPARClike} operates on received signals of the form
\begin{equation}
\yv = \mathbf{A} \mathbf{D} \sv + \zv
\end{equation}
where $\sv = \sv(1) \cdots \sv(L)$ is partitioned into $L$ sections.
Matrix $\mathbf{D}$ is diagonal with equal non-negative entries within each section; it accounts for the power allocated to every section.
In the spirit of AMP for sparse regression codes~\cite{joseph2013fast,CIT-092,rush2017capacity}, the idea underlying the decoding of the inner code is to create a two-step iterative process to recover the sparse vector $\sv$.
This algorithm alternates between the following two equations for $t=0,1,\ldots$:
\begin{align}
&\zv^{(t)} = \yv - \mathbf{A} \mathbf{D} \sv^{(t)} + \frac{\zv^{(t-1)}}{\tau_{t-1}^2}
\left( \Ka P - \left\| \mathbf{D} \sv^{(t)} \right\|^2 \right)
\label{equation:AMP} \\
&\sv^{(t+1)}(\ell) = \eta_{\ell}^{(t)} \left( \sv^{(t)}, \zv^{(t)} \right)
\label{equation:denoiser}
\end{align}
where $\zv^{(-1)} = \mathbf{0}$, $\sv^{(0)} = \mathbf{0}$, $\tau_t^2 = \frac{\| \zv^{(t)} \|^2}{n}~\forall~t \ge 0$~\cite{greig2017techniques}.
The denoiser of \eqref{equation:denoiser} seeks to produce an estimate for $\sv^{(t+1)}$, block-wise, based on the effective observation $\mathbf{D} \sv^{(t)} + \mathbf{A}^{*} \zv^{(t)}$.
The authors in \cite{Giuseppe} demonstrate that AMP can be adapted to the application scenario at hand, although this is only possible after addressing several technical challenges rooted in the block sparse structure of the problem.
We briefly explain these challenges
and emphasize places where our implementation differs from theirs.

To gain a better understanding of AMP decoding, we begin by looking at \eqref{equation:denoiser}, in which $\sv^{(t+1)}$ is essentially generated based on the statistic $\rv^{(t)} = \mathbf{D} \sv^{(t)} + \mathbf{A}^{*} \zv^{(t)}$.
The structure of the AMP denoiser is predicated on the presumption that $\rv^{(t)}$ should be asymptotically distributed ($n \rightarrow \infty$) as $\mathbf{D} \sv + \overline{\tau}_t Z$, where $\overline{\tau}_t$ is asymptotically related to $\tau_t$ and $Z$ is an i.i.d.\ $\mathcal{N}(0,1)$ random vector, independent of $\sv$.

In the original AMP for SPARCs implementation~\cite{rush2017capacity}, the denoiser is applied independently to every section.
Furthermore, given the aforementioned Gaussian structure, a natural choice for an update in this situation is the conditional mean estimator $\sv^{(t+1)} = \mathbb{E} \left[ \sv | \mathbf{D} \sv + \tau_t Z = r \right]$.
While this approach works adequately for one-sparse sections, it does not transfer directly to the unsourced MAC problem.
A suitable approximation to the optimal solution, at least for select parameters, is based on the marginal posterior mean estimate (PME) of Fengler et al.~\cite{Giuseppe}.
This quantity can be expressed as
\begin{align} 
\hat{s}_{\ell}^{\mathrm{OR}} \left( r, \tau \right)
&= \frac{q e^{ - { \left( r - d_{\ell} \right)^2}/{2 \tau^2} }}
{(1-q) e^{ -{r^2}/{2 \tau^2} }
+ q e^{ - { \left( r - d_{\ell} \right)^2}/{2 \tau^2} }}  \label{equation:OriginalPME}
\end{align}
where 
$q = \Pr (s = 1) = 1-\left(1-2^{-v_{\ell}}\right)^{\Ka}$ is a constant.
The overall estimate is given by
\begin{equation*}
\hat{\sv}^{\mathrm{OR}} \left( \rv, \tau \right)
= \hat{\sv}_{1}^{\mathrm{OR}} \left( \rv(1), \tau \right) \cdots
\hat{\sv}_{L}^{\mathrm{OR}} \left( \rv(L), \tau \right)
\end{equation*}
where the $k$th element of $\hat{\sv}_{\ell}^{\mathrm{OR}} \left( \rv(\ell), \tau \right)$ is obtained by applying $\hat{s}_{\ell}^{\mathrm{OR}} ( \cdot, \tau )$ using the $k$th element of $\rv(\ell)$ as its first argument.
The insight behind this approach is that, instead of estimating signal $\sv$ directly, it suffices to infer its support using the marginal PME.
This low-complexity strategy offers good empirical performance when combined with AMP, as reported in~\cite{Giuseppe}.
Furthermore, this approach for CCS-AMP is aligned with the original AMP for SPARCs algorithm in that the denoiser is applied section by section.

Yet, given the presence of a tree code, the probabilities on the elements of $\sv(\ell)$ depend on the probability distributions over (graph) neighboring sections.
This viewpoint is a departure from AMP for SPARCs in the context of the unsourced MAC~\cite{Giuseppe}.
The motivation behind this connection is that, although $\vv$ is a vector of length $2^v$, the tree encoding process forces it to lie in a random set of cardinality $2^w$.
Likewise, the tree code confines every section to take value in a potentially much smaller subset when conditioned on neighboring sections.
This is in stark contrast with the treatment in~\cite{Giuseppe} where $\vv(\ell)$ is implicitly assumed uniform over all $2^{v_{\ell}}$ possibilities, with uninformative priors of the form $q = 1-\left(1-2^{-v_{\ell}}\right)^{\Ka}$ applied to \eqref{equation:OriginalPME}.
The hope is then to integrate this knowledge into the AMP algorithm to take advantage of this embedded structural property.
To do so, we must delve deeper into the structure of the current problem.
The idea is to compute the marginal PME in a manner analogous to \eqref{equation:OriginalPME}, but to incorporate the prior probability on $\sv(\ell)$ inherited from $\{ \sv(k) : k \in [1:L] \setminus \ell \}$.
This warrants a small modification to the marginal PME, enhancing the function with an argument that enable passing prior probabilities.
The posterior mean estimator (PME) $\hat{s}_{\ell}^{\mathrm{OR}} \left( q, r, \tau \right)$ of the decoupled model takes the same form as \eqref{equation:OriginalPME}, albeit prior probability $q$ is exposed as an argument.
The information contained in the tree code can then be incorporated into the creation of a marginal PME.

Having established a notation for the marginal PME that takes into account the information from neighboring sections, we can rewrite the denoiser of \eqref{equation:denoiser} component-wise as
\begin{equation*}
\begin{split}
\left( \eta_{\ell}^{(t)} \left( \sv^{(t)}, \zv^{(t)} \right) \right)_k \! \!
= \hat{s}_{\ell}^{\mathrm{OR}} \left( {q}_{\ell}(\vv(\ell)), \left( \mathbf{D} \sv^{(t)} + \mathbf{A}^{*} \zv^{(t)} \right)_k, \tau_t \right)
\end{split}
\end{equation*}
where $(\cdot)_k$ denotes the entry of the argument vector at location $k = [\vv(\ell)]_2$ and ${q}_{\ell}(\vv(\ell)) = 1-\left(1-\tilde{q}_{\ell}(\vv(\ell))\right)^{\Ka}$.
The quantity $\tilde{q}_{\ell}(\vv(\ell))$ is implicitly a function of $\sv^{(t)}$ and it carries the same meaning as the probability developed in Section~\ref{subsection:TreeDecoding}.
That is, it is derived leveraging the structure of the tree code, computing priors based on the (truncated) dependency tree.

\section{Simulation Results and Discussion}
\label{section:SimulationResults}

Numerical results demonstrate the performance improvement offered by the proposed enhanced decoder over the one employed in \cite{Giuseppe} for  $\Ka \in [10:300]$. The size of the payload is $w = 128$ bits. The total number of channel uses is $n = 38400$. The target per-user probability of error is $P_e = 0.05$.
The length of each sub-block is set to $16$ bits, i.e., $v_{\ell} = 16~\forall~\ell \in [1:L]$.

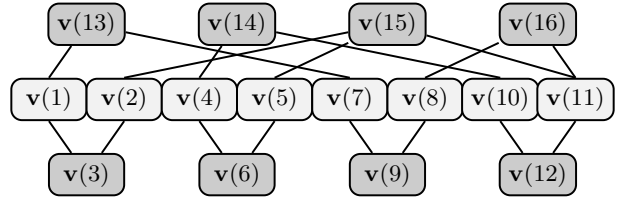
\begin{figure}[t!]
  \centering
  \begin{tikzpicture}[
  font=\small, >=stealth', line width=0.75pt,
  infobits/.style={rectangle, minimum height=3mm, minimum width=10mm, draw=black, fill=gray!10, rounded corners},
  paritybits/.style={rectangle, minimum height=3mm, minimum width=10mm, draw=black, fill=gray!40, rounded corners}
]

\node[infobits] (vb1) at (1,0) {$\vv(1)$};
\node[infobits] (vb2) at (2,0) {$\vv(2)$};
\node[infobits] (vb4) at (3,0) {$\vv(4)$};
\node[infobits] (vb5) at (4,0) {$\vv(5)$};
\node[infobits] (vb7) at (5,0) {$\vv(7)$};
\node[infobits] (vb8) at (6,0) {$\vv(8)$};
\node[infobits] (vb10) at (7,0) {$\vv(10)$};
\node[infobits] (vb11) at (8,0) {$\vv(11)$};

\node[paritybits] (vp3) at (1.5,-1) {$\vv(3)$};
\node[paritybits] (vp6) at (3.5,-1) {$\vv(6)$};
\node[paritybits] (vp9) at (5.5,-1) {$\vv(9)$};
\node[paritybits] (vp12) at (7.5,-1) {$\vv(12)$};
\node[paritybits] (vp13) at (1.5,1) {$\vv(13)$};
\node[paritybits] (vp14) at (3.5,1) {$\vv(14)$};
\node[paritybits] (vp15) at (5.5,1) {$\vv(15)$};
\node[paritybits] (vp16) at (7.5,1) {$\vv(16)$};

\draw  (vb1.south) edge (vp3);
\draw  (vb2.south) edge (vp3);
\draw  (vb4.south) edge (vp6);
\draw  (vb5.south) edge (vp6);
\draw  (vb7.south) edge (vp9);
\draw  (vb8.south) edge (vp9);
\draw  (vb10.south) edge (vp12);
\draw  (vb11.south) edge (vp12);

\draw  (vb1.north) edge (vp13);
\draw  (vb7.north) edge (vp13);
\draw  (vb4.north) edge (vp14);
\draw  (vb10.north) edge (vp14);
\draw  (vb2.north) edge (vp15);
\draw  (vb5.north) edge (vp15);
\draw  (vb11.north) edge (vp15);
\draw  (vb8.north) edge (vp16);
\draw  (vb11.north) edge (vp16);

\end{tikzpicture}
  \caption{This graph shows the connections between information and parity blocks used in the numerical simulations.
  Shaded blocks represent parity sections.
  }
  \label{figure:architecture}
\end{figure}
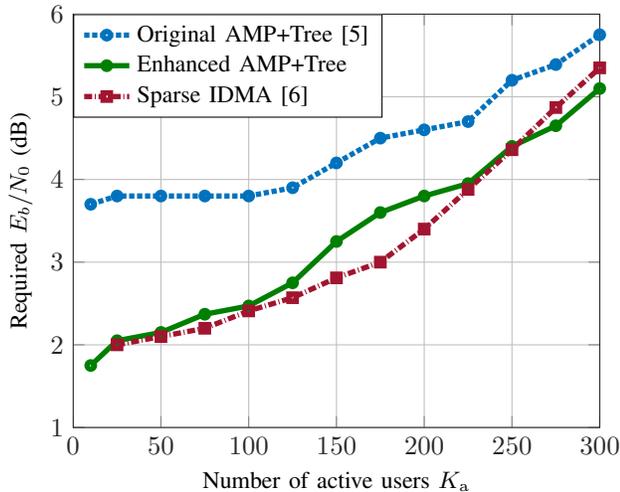
\begin{figure}[h]
\centerline{\begin{tikzpicture}
\definecolor{mycolor1}{rgb}{0.63529,0.07843,0.18431}%
\definecolor{mycolor2}{rgb}{0.00000,0.44706,0.74118}%
\definecolor{mycolor3}{rgb}{0.00000,0.49804,0.00000}%
\definecolor{mycolor4}{rgb}{0.87059,0.49020,0.00000}%
\definecolor{mycolor5}{rgb}{0.00000,0.44700,0.74100}%
\definecolor{mycolor6}{rgb}{0.74902,0.00000,0.74902}%

\begin{axis}[%
font=\small,
width=7cm,
height=5.5cm,
scale only axis,
every outer x axis line/.append style={white!15!black},
every x tick label/.append style={font=\color{white!15!black}},
xmin=0,
xmax=300,
xtick = {0,50,100,...,300},
xlabel={Number of active users $\Ka$},
xmajorgrids,
every outer y axis line/.append style={white!15!black},
every y tick label/.append style={font=\color{white!15!black}},
ymin=1,
ymax=6,
ytick = {1,...,6},
ylabel={Required $E_b/N_0$ (dB)},
ymajorgrids,
legend style={at={(0,1)},anchor=north west, draw=black,fill=white,legend cell align=left}
]

\addplot [color=mycolor2,densely dotted,line width=2.0pt,mark size=1.4pt,mark=o, mark options={solid}]
  table[row sep=crcr]{10 3.7\\
25	3.8\\
50	3.8\\
75	3.8\\
100	3.8\\
125	3.9\\
150	4.2\\
175 4.5\\
200 4.6\\
225 4.7\\
250 5.2\\
275 5.39\\
300 5.75\\
};
\addlegendentry{Original AMP+Tree \cite{Giuseppe}};

\addplot [color=mycolor3,solid,line width=2.0pt,mark size=1.4pt,mark=o,mark options={solid}]
  table[row sep=crcr]{10 1.75\\
  25  2.05\\
50	2.15\\
75	2.37\\
100	2.47\\
125	2.75\\
150	3.25\\
175 3.6\\
200 3.8\\
225 3.95\\
250 4.4\\
275 4.65\\
300 5.1\\
};
\addlegendentry{Enhanced AMP+Tree};

\addplot [color=mycolor1,densely dashdotted,line width=2.0pt,mark size=1.4pt,mark=square,mark options={solid}]
  table[row sep=crcr]{
  25  2\\
50	2.1\\
75	2.2\\
100	2.41\\
125	2.57\\
150	2.81\\
175	3\\
200 3.4\\
225 3.88\\
250 4.36\\
275 4.87\\
300 5.35\\
};
\addlegendentry{Sparse IDMA \cite{pradhan2019joint}};

\end{axis}

\end{tikzpicture}%
}
  \caption{The figure compares the performance of the proposed scheme with existing schemes.}
  \label{fig:resultsISIT}
\end{figure}

Figure~\ref{figure:architecture} shows the explicit graph used for these simulations for $\Ka < 200$.
When $\Ka \ge 200$, we add two additional parity sections to reduce the probability of this decoder producing a list of size greater than $\Ka$.
While this architecture seems to offer good performance for the parameters of interest, we remark that other frameworks are worthy of investigation. 
Sensing matrix $\mathbf{A} \in \mathbb{R}^{n \times m}$ is formed by picking $n$ rows uniformly at random from a Hadamard matrix of dimension $m \times m$.
If the decoder produces a list of size greater than $\Ka$, we choose the $\Ka$ messages with largest likelihoods.
The topmost curve in Fig.~\ref{fig:resultsISIT} demonstrates the performance of scheme presented in \cite{Giuseppe} which uses uninformative priors.
The performance improvement associated with CCS-AMP that uses informed priors is captured by the second curve.
It can be seen that the enhanced decoder outperforms the original decoder in \cite{Giuseppe} for all values of $\Ka$ and the gain is more pronounced for small values of $\Ka$.
The bottom curve, which is based on sparse-IDMA represents the state-of-the-art for $\Ka \ge 250$.
Our proposed scheme outperforms this current best known solution for $\Ka \ge 250$.
Preliminary results for CCS-AMP are very encouraging.
Although the coding architecture has not been fully optimized, its performance is excellent and complexity remains manageable.





\bibliographystyle{IEEEbib}
\bibliography{IEEEabrv,MACcollison}

\end{document}